# Nested economies of scale in global city mass

Kangning Huang[1,2*], Mingzhen Lu[3*]

**Author affiliations:**
[1]Shanghai Key Laboratory of Urban Design and Urban Science, NYU Shanghai, Shanghai 200124, China
[2]Division of Arts and Sciences, NYU Shanghai, Shanghai 200124, China
[3]Department of Environmental Studies, New York University, New York, NY 10003, USA

*Both authors contributed equally

**E-mail addresses (ORCID):**

K. Huang: kangning.huang@nyu.edu (0000-0001-6877-9442);
M. Lu: mingzhen.lu@nyu.edu   (0000-0002-8707-8745);

**Corresponding authors**:
Kangning Huang,
Division of Arts and Sciences,
New York University Shanghai,
567 West Yangsi Road, Shanghai, China
Email: kangning.huang@nyu.edu

Mingzhen Lu,
Department of Environmental Studies
New York University,
59 Washington Square East, New York, NY 10003, USA
Email: mingzhen.lu@nyu.edu

**Abstract**

A longstanding puzzle in urban science is whether there's an intrinsic match between human populations and the mass of their built environments. Previous findings have revealed various urban properties scaling nonlinearly with population, yet existing models of city built mass are still dominated by per-capita linear thinking. Our analysis of >3,000 cities globally reveals universal sublinear scaling of city mass with population at both the city ($\beta$=0.90) and neighborhood levels ($\delta$=0.75). This means that larger cities and denser neighborhoods achieve economies of scale with less per-capita built mass. Our theoretical framework further shows that city-level scaling emerges naturally from within-city disparities. This multi-scale understanding redefines "over-built" and "under-built" conditions as deviations from expected scaling patterns, implying either excessive environmental impacts or inadequate living standards. Effective urban policy thus requires moving beyond simple per-capita assumptions, adopting scale-adjusted metrics and managing cities as nested, complex systems.

**Main text**

A key feature of human life is the extensive transformation of Earth's materials into built structures, establishing the biophysical foundation of modern civilization[1]. This capacity to harness external mass has progressively intensified since the Stone Age, evolving from simple makeshift shelters in early human settlement[2] into today's complex city built environment[3], where individuals find comfort and protection within individual buildings, enjoy local conveniences like grocery shopping and schooling within vibrant neighborhoods, and traverse city-wide transportation networks for professional collaborations and cultural exchanges[4].

Earth-scale data from recent breakthrough studies[1,5,6] have allowed us to quantify this global intensification of human material use. In particular, global per-capita built mass has grown exponentially, with an approximate annual growth rate of 1.8% over the past 120+ years (Fig.1a, Methods). Expressed dimensionlessly relative to average human body mass (~50 kg, Methods), each person now corresponds to about 4,090× their body mass in built structure (205 tonnes) (Fig.1a). This material footprint is projected to continue increasing with rising living standards worldwide[7].

Alongside the growing per-capita material footprint, the UN projects a global urban expansion of over 2 billion more residents by the mid-century[8]. Combined, this dual-growth is projected to drive annual material use for urban construction and maintenance from 40 billion tonnes in 2010 to about 90 billion tonnes by 2050[9].

However, this projection has two key limitations: 1) It relies on linear models, implicitly assuming that per-capita material consumption depends solely on living standards and affluence[7], independent of city size (*e.g.*, doubling population will double consumption). But, recent advances [10–12] in urban scaling research challenges this assumption, showing that many urban properties scale nonlinearly with city size (*e.g.*, doubling population more than doubles economic output), making per-capita predictions questionable. 2) It lacks spatial detail, overlooking significant variations in infrastructure within cities[13–15]. This limits insight into whether nonlinear scaling holds across spatial scales, from cities to neighborhoods. These gaps—linear assumptions and low spatial resolution—hinder accurate predictions of material use and infrastructure needs across diverse urban areas, risking misguided sustainability and climate policies.

Here we map, for the first time, the global mass distribution of the human built stock (human-supporting buildings, roads, other infrastructures) by integrating high-resolution mappings of buildings[16–18], mobility infrastructure[19], and pavement[20] in >3,000 cities[21] (Fig.1b, Supplementary Fig.1,4-6, Methods) with region/type-specific material intensities[22,23].

Our analysis offers an in-depth look of the built-structure *vs.* human match in urban settings. The total built mass of cities worldwide is a staggering 324 gigatonne, equivalent to every inch of global urban impervious area[20] (~ 0.3M km$^2$) submerged under half meter deep of solid concrete (height of a coffee table). All cities taken together, each person is supported by built structure—buildings and mobility infrastructure such as paved roads and parking lots—that is >5,000 times their own mass while only enjoying vegetation 64 times of their mass (Fig.1c). This finding suggests that albeit more densely built, urban built structure vs. human match is on par with the global average (~4,000×), but urban systems have dramatically less vegetation per person (64× vs. 2,768× global average).

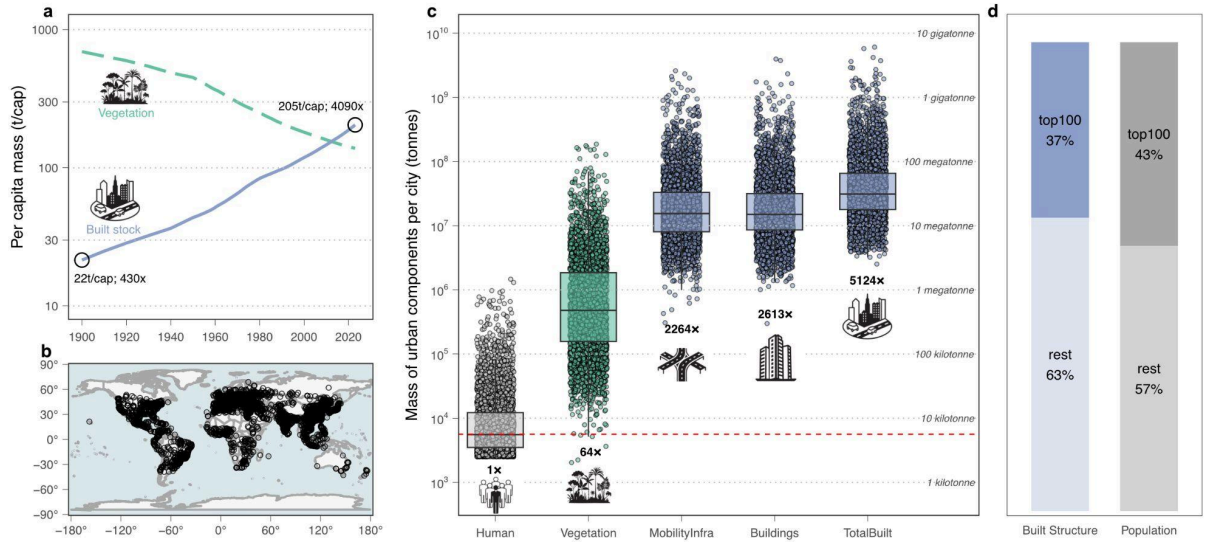

**Figure 1| Global mass distribution of the built stock in relation to vegetation and humans.**
(**a**), Per-capita built stock increased exponentially from 22t/cap in 1900 to 205t/cap today, a 1.8% annual growth. In contrast, per-capita vegetation biomass declined from 694t/cap to 138t/cap, a 1.3% annual decrease. Note that the y axis is logarithmic scale. (**b**) Map of 3593 cities included in our analysis with a size cutoff of 50,000 residents (Methods). (**c**) City-level mass of the built environment is ranked according to the median value: Human (gray), vegetation biomass (green), mobility infrastructure (blue), buildings (blue), total built stock (blue). The red dashed line (median city-level human mass) serves as a reference threshold across categories. Scale indicators below each category (*e.g.*, 64×) denote the average multiplicative difference relative to human mass (average human body mass estimated to be 50kg), highlighting the magnitude of resources required to support human living. (**d**) The world's biggest 100 cities account for 37% of all mass and 43% of urban population. In each box plot, the lower and upper bounds of the whiskers denote minima and maxima, the center line denotes the median, and the lower and upper bounds of the boxes represent the 25% and 75% quantiles, respectively.

Looking beyond per-cap averages, the distribution of city mass is highly right-skewed (Fig.1d): the world's heaviest 100 cities—less than 3% of all cities—alone account for ~37% of all urban mass, equivalent to the mass of the smallest 3,314 cities combined (~90% of cities). The distribution of population is similarly more concentrated in big cities: the top 100 populous cities alone host 43% of the global urban population. Curiously, the top 100 most populous cities in the world host more share of population with disproportionately less built mass, resulting in an approximately 32% lower per capita built mass compared to the rest of the world (172 *vs.* 252 t/cap).

This notable difference suggests that the per-capita ratio may not be scale-invariant and could potentially decline as cities grow larger. In the next section, we analyze the nonlinear dependence of built mass as a function of city size.

### It takes size to share: universal across-city economies of scale

We found a consistent sublinear scaling relationship between urban built mass ($M$) and city size measured by population ($N$) across major countries of the world (Fig.2a; $M = M_0 N^\beta$), pointing to economies of scale in material use as cities get bigger. Most countries have a scaling slope ($\beta$) around 0.88 (<1, sublinear) despite the wide range of economic development level, technology, culture of building style, climatic conditions, and abundance of natural resources included in our global datasets. The distribution of scaling slopes (Fig.2a, left inset) suggests that there's a intrinsic central tendency of the urban system around 0.88, and country-scale variations generate deviations from this central tendency in a Gaussian manner.

Note that this scaling slope (0.88) is less sublinear than the slope for road area derived from urban scaling theory (0.83)[10], which did not account for the varying per-area material mass. This difference arises because the most heavily used roads require disproportionately more material per unit area (*e.g.*, thicker pavement layers)[22] to withstand higher traffic loads and ensure structural durability, resulting in a built mass that grows less efficiently with population than mere surface area alone.

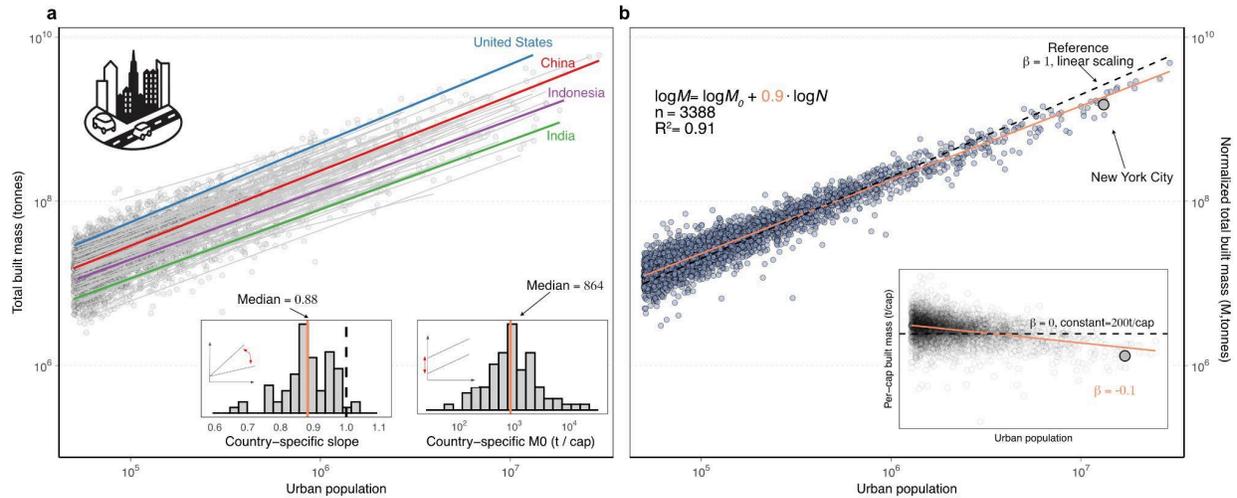

**Figure 2| Sublinear urban scaling of built mass and variations are universal. (a)** Log-log scatter plot of total city mass (tonnes) scales sublinearly with population across cities (gray dot) from 74 countries, indicating economies of scale in urban infrastructure materials. A country is included if they have more than 5 cities. We colored 4 countries to highlight similarity of slopes and variations of intercept. All other countries are denoted by gray lines (see Supplementary Fig.2 for regressions of all countries and Supplementary Fig.3 for distribution of goodness of fit). Inset on the left describes the distribution of country-specific scaling slope (*n*=74, median=0.88), while inset on the right describes the distribution of country-specific intercept $M_0$ (*n*=74, median = 864 t/cap$^\beta$). **(b)** After normalizing cities based on their country-specific intercepts (Methods, linear mixed effects model), the solid orange line shows the best-fit power-law scaling relationship (exponent $\beta = 0.9$, $R^2 = 0.91$, $n = 3,388$). The dashed black line represents the linear expectation if city mass were directly proportional to population size based on per-capita average ($\beta = 1$). The inset highlights the divergence between these two lines expressed as per-capita built mass, underscoring the inadequacy of linear estimates. By definition of per-capita metric, $\beta=0.9-1 = -0.1$. Each blue dot represents a unique city based on our definition (Supplementary Fig.1); we highlighted New York City as an example.

In contrast to the Gaussian distribution of *slope*, the *intercept* of the scaling relationship ($M_0$) has a right-skewed distribution: some countries have extremely large values while the majority of countries are much less materially developed and have small values. This country-scale variation in intercept seems to follow a log-normal distribution (Fig.2a, right inset, median = 864 t/cap$^\beta$), reflecting a high degree of disparity in baseline built mass across countries, influenced by factors such as income levels, construction practices, and historical development patterns.

Given the centralized distribution of scaling slope, we were able to normalize our city data by removing country-specific variations in baseline to characterize the general relationship between built mass and city size (Methods). We found a universal sublinear scaling law across all cities in our dataset (Fig.2b, $\beta = 0.9$, 95% CI = (0.89,0.91), $r^2 = 0.91$, n= 3,588, P<0.001). A city with twice the population of another city only needs 84% more mass, a 16% efficiency gain for every doubling of population size.

As a consequence, per-capita built mass declines as cities get bigger (by definition of "per-capita", $\beta$ = 0.9-1=-0.1, Fig.2b inset). A city that is 10-times bigger is only about ¾ in per-capita mass compared to the smaller city, proving that per-capita mass is not a good metric for comparing cities of different sizes (see contrast between dashed black and orange solid lines in Fig.2b).

The scale-induced efficiencies can largely be attributed to the shared use of infrastructure in larger urban areas. As cities grow, they facilitate the communal utilization of resources such as roads, utilities, and public spaces, leading to economies of scale that reduce the per capita built mass. This phenomenon aligns with the concept of urbanization economies[24,25], where the concentration of population and activities in urban centers enhances productivity and resource efficiency through shared infrastructure and services. These findings extend previous studies on urban scaling[10,26,27], which demonstrated that the lengths of infrastructure—such as roads, cables, and pipes—scale sublinearly with population due to the increasing efficiency of spatial connectivity in larger urban systems.

However, previous studies have largely focused on the geometry[27–29] (*e.g.*, length or area) of infrastructure rather than its material embodiment. Our analysis introduces a new dimension: the total mass of the built environment, which offers a more direct account of the material and energetic investments embedded in infrastructure.

While these results underscore the universality of sublinear scaling at the city level, they prompt further inquiry into whether similar patterns hold within cities themselves. Despite recognition of cities displaying self-similar features at various spatial scales, previous efforts of urban scaling analysis have largely assumed cities to be homogenous. According to the homogeneity assumption, cities are internally uniform, meaning that their neighborhoods are similar to each other in terms of population density and built environment, rather than exhibiting the diverse and uneven patterns that actually characterize real urban areas. In the next section, we challenge this homogeneity assumption by examining how built mass scales with population at the level of neighborhoods.

**Economies of scale at the neighborhood level**

Neighborhoods are fundamental units of human settlement, shaping daily life, economic opportunity, and urban form from ancient times to the present. While urban sociologists have long highlighted their importance[30,31], neighborhoods have rarely been included in urban scaling analyses, largely because most scaling studies have focused on whole cities as the unit of analysis and lacked the high-resolution spatial data needed to analyze intra-city variation. In this paper, we define neighborhoods using a global hexagonal grid (H3, ~3.7km radius; Methods)[32] at a spatial scale that captures the extent of day-to-day human activity.

Across a global dataset of urban neighborhoods, we observe that built material mass increases sublinearly with population in both across-city and across-neighborhood comparisons. Importantly, because each neighborhood polygon in our sample has nearly the same land area, its "population" can be considered a measure of density—unlike whole-city population size, which combines density and total area. Using a multilevel fixed-effects model (see Methods), with separate intercepts by city and country but a common slope, we estimate a markedly lower scaling exponent within cities ($\delta$ = 0.75; red line in Fig.3a) than across entire cities ($\beta$ = 0.90; orange line). In other words, local density effects within a city generate even more pronounced economies of scale than aggregate city-to-city comparisons.

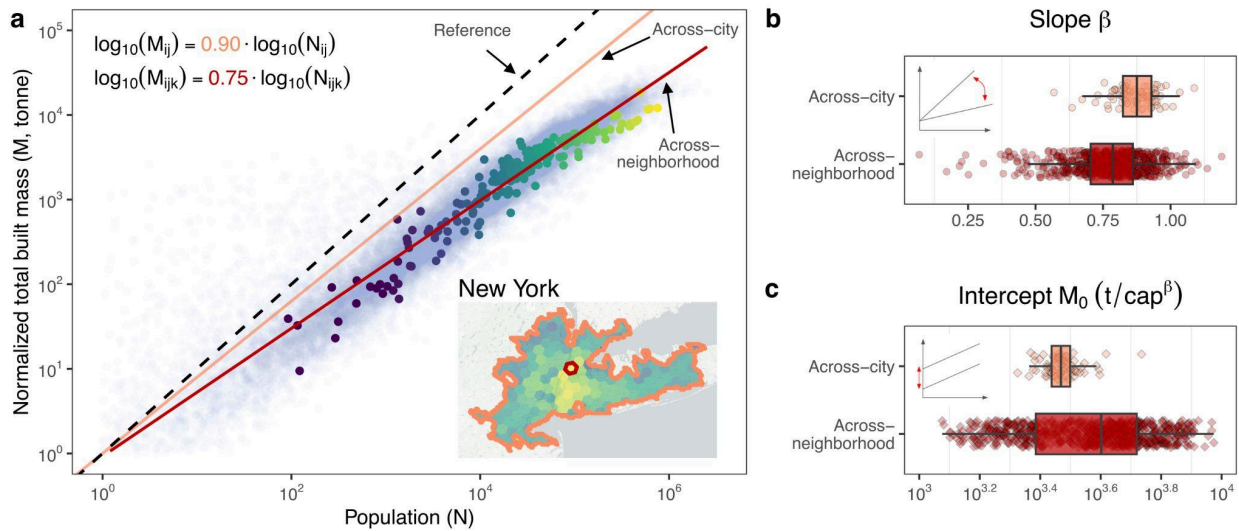

**Figure 3| Sublinear scaling of built mass across neighborhoods.** (**a**) Log-log scatter plot of the normalized total built mass *M* (in tonnes) versus population *N* for the *k*-th neighborhood (in the *j*-th city and the *i*-th country), with regression lines showing the across-neighborhood scaling relationship (red, slope $\delta = 0.75$; n=34,045; $R^2$=0.85) and the across-city relationship (orange, slope $\beta = 0.9$). The dashed black line indicates linear scaling (slope = 1). Colored points highlight neighborhoods in New York, with an inset hex map showing their spatial distribution of population. Boxplots of the fitted slopes (**b**) and intercepts (**c**) from these regressions compare the distributions of across-city fits within countries (orange) and across-neighborhood fits within cities (red), showing larger variations in the across-neighborhood regressions.

At two different scales–neighborhoods within cities (red) and cities within countries (orange)--the fitted slopes (Fig.3b) and intercepts (Fig.3c) reveal distinct scaling relationships . Finer-scale analyses display greater variability, with the distribution of neighborhood-level slopes not only broader but also shifted toward lower values (i.e., more sublinear). The across-neighborhood intercepts (representing city-specific baseline) also span a wider range than that of across-city intercepts (representing country-specific baseline), suggesting that cities are far more diverse than countries in terms of land-use intensity and the clustering of infrastructure.

Our findings expand upon and unify several strands of recent research. Prior studies[14,22] have found lower per capita mobility infrastructure mass in denser urban areas. Other research[13,33] similarly indicates that total built mass per capita is lower in denser urban settings but without systematically quantifying these relationships. Sublinear scaling patterns for building volumes have been identified within 10 individual Chinese cities [34], but this study is geographically limited, raising the question of global generalizability.

Our analysis addresses this gap by demonstrating a universal sublinear scaling trend in total built mass at the neighborhood level across thousands of cities worldwide. Moreover, by explicitly contrasting within-city and across-city patterns, we show that stronger economies of scale emerge at finer spatial resolutions—underscoring the importance of multi-scalar analysis in urban metabolism research. This raises a key theoretical question: why does the scaling exponent increase when moving from local ($\delta$ =0.75) to aggregate scales ($\beta$ =0.9), and what mechanisms can bridge these levels?

**Emergence of across-city sublinear scaling from across-neighborhood disparity**
Bigger cities tend to develop denser neighborhoods because agglomeration effects drive up land values, creating strong incentives to build upward and cluster people together to maximize the benefits. If all residents in bigger cities were concentrated in denser neighborhoods—leveraging those steep economies of scale—the city-level scaling could approach 0.75. But in reality, even large cities mix low- and high-density areas, so not all population benefits equally from density efficiencies, pushing the aggregate $\beta$ higher. The unevenness of within-city population distribution likely determines how much neighborhood-level savings "trickle up" to shape city-wide patterns.

We thus hypothesize that the degree of population heterogeneity is the key that bridges scaling across the two nested spatial scales: across cities ($\beta = 0.90$) and across neighborhoods ($\delta = 0.75$). Existing research suggests that neighborhood populations often conform to Zipf-like, heavy-tailed distributions, where a small number of dense neighborhoods house disproportionately larger amounts of residents[15,35,36]. Given the sublinear scaling of built mass at the neighborhood level, these dense areas benefit from enhanced economies of scale, reducing the per capita built mass at the city-wide level.

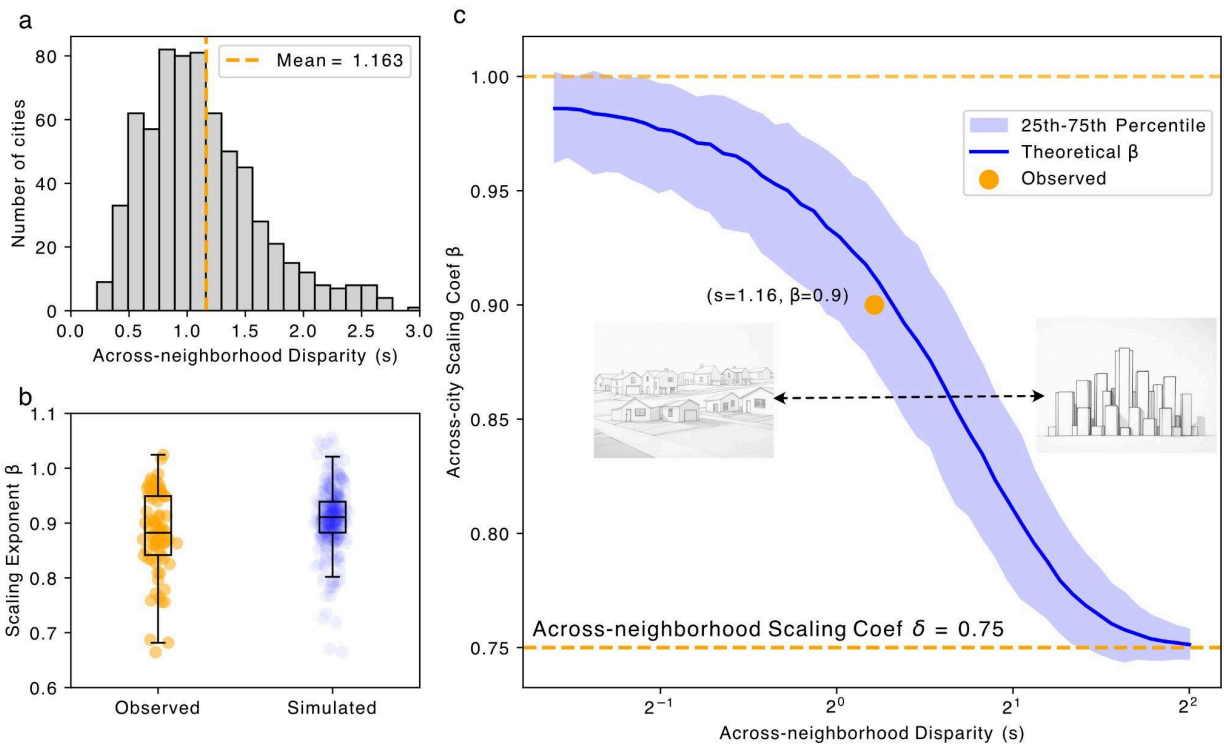

**Figure 4| Emergence of sublinear across-city scaling from across-neighborhood disparity.** The disparity parameter $s$ (Zipf's exponent) captures how uneven population is distributed across neighborhoods—larger $s$ indicates greater disparity (a few dense neighborhoods dominate), while smaller $s$ reflects more uniform distributions. (**A**) Probability distribution of $s$ across cities, with the mean ($s = 1.16$) marked by a dashed line. (**B**) Boxplots comparing observed city-level scaling exponents $\beta$ (orange) with those predicted by theoretical simulations (blue), showing strong agreement. Simulated cities assume Zipf-like neighborhood population distributions and a fixed within-city scaling exponent ($\delta=0.75$). In each box plot, the lower and upper bounds of the whiskers denote minima and maxima, the center line denotes the median, and the lower and upper bounds of the boxes represent the 25% and 75% quantiles, respectively. (**C**) Theoretical relationship between $s$ and $\beta$ (solid blue line): a system of more homogenous cities (illustration showing cities with uniform single family housing districts, small $s$) approaches linear scaling (upper bound, $\beta \approx 1$), while a system of more heterogeneous cities (illustration showing cities with buildings of varying heights, large $s$) converges toward the within-city exponent (lower bound, $\delta=0.75$). The empirical means ($s=1.16$, $\beta=0.90$, orange point) aligns with this trend.

To test this hypothesis, we developed a theoretical model comprising simulated cities with neighborhoods whose populations follow a Zipf-like distribution described by an across-neighborhood disparity parameter $s$ (see Methods). In this setup, the population of the $i$-th ranked neighborhood is $1/i^s$ of the largest neighborhood: populations are more concentrated in denser neighborhoods in cities with larger $s$. Neighborhood-level built mass was assigned using the empirical scaling exponent ($\delta = 0.75$), then aggregated to obtain total city-level built mass ($M$) and population ($N$). We estimated the across-city scaling exponent $\beta$ by regressing $\log(M)$ against $\log(N)$ across these simulated cities. We introduced two random processes to capture the variability: 1) we drew a random $s$ parameter from a normal distribution, centered at the given $s$ with a standard deviation (0.59) estimated from the observed distribution (Fig.4a); 2) we grouped the cities randomly into "countries" and conducted separate regressions, producing a distribution of theoretical $\beta$ values.

We found that the observed distribution of disparity parameter $s$ is right-skewed (Fig.4a), which suggests that a small number of cities have very heterogeneous neighborhoods (*i.e.*, coexistence of very dense and very sparse neighborhoods). Based on the observed distribution of $s$ (mean=1.16, std=0.59), we simulated across-city scaling exponents ($\beta$), which strongly aligns with observed exponents (Fig.4b). By sweeping various values of $s$, our simulations further elucidate the theoretical relationship between neighborhood disparity $s$ and city-level scaling exponent $\beta$, depicting a smoothly declining curve from nearly linear ($\beta \approx 1.0$) at low disparity to strongly sublinear ($\beta$ approaching $\delta= 0.75$) at high disparity (Fig.4c). Our mathematical derivation (Methods) further illustrates that across-city scaling exponent $\beta$ is intrinsically bounded by two limits ($\delta \leq \beta \leq 1$), explaining the increase of exponent when moving from local (neighborhood) to the aggregate (city).

Our empirical analysis and theoretical simulations indicate that within-city density disparities are a critical mechanism underpinning economies of scale in built material use. Denser neighborhoods—often housing a disproportionate share of the urban population—require less built mass per capita, thereby driving the sublinear scaling observed at the city level. Classical urban economic models[37–39] provide a well-established explanation for the emergence of these disparities. In monocentric cities, proximity to the central business district raises land values and incentivizes denser development near the core. In polycentric cities, multiple subcenters generate additional peaks of accessibility and land value, sustaining density gradients as households and firms cluster around these nodes[40].

However, zoning regulations and density constraints can suppress this natural tendency toward concentrated development[41]. Restrictions such as maximum floor-area ratios, height limits, or exclusionary zoning limit high-density construction in areas with strong market demand, effectively flattening the urban density gradient (artificially decreasing $s$ and increasing $\beta$)[42]. As a result, such anti-density interventions may unintentionally nullify the material efficiency gains associated with urban density variation—undermining the very mechanism that enables cities to use built materials more efficiently.

**Cities as nested complex systems of built structure**
Our analysis indicates that conventional per-capita projections systematically overestimate the required built mass in larger cities and denser neighborhoods. According to the latest IPCC assessment report[43,44], cement and steel—critical materials primarily used for built mass—account for ~5.2 GtCO$_2$-eq/yr of global emissions. From 2018 to 2050, the cumulative buildings, infrastructure and machinery-related emissions are projected[7] to reach between 1,300 and 2,100 Gt CO$_2$-eq, driven mainly by increasing material stocks. Yet these projections treat every future urban resident as if they would require the same amount of material irrespective of the size or form of the city they live in.

Neglecting size-dependent nonlinear scaling can lead to substantial mis-estimations in material demand—an issue made clear when re-examining a recent UN-Habitat projection[45] through the lens of our findings. This projection shows that, in low-income countries, the population living in megacities ($N>$ 5 millions) will rise from 18 million today to about 140 million by 2070. In those countries, the projected future population living in megacities will be roughly equal to the population (~160M) that now inhabit small cities ($N<$250k). Under the across-city scaling relationship revealed here ($\beta = 0.9$), multiplying city population size by twenty (250k → 5M) raises per-capita material demand by only $20^\beta \approx 14.8$, not twenty-fold. This means that accommodating the same number of people in megacities, instead of small cities, reduces the per-capita material footprint by roughly 26% (1 − 14.8/20 ≈ 0.26). Therefore, policies that cap city size could erase an important material-efficiency gain.

Within cities we find a steeper sub-linear scaling of built mass with population density ($\delta = 0.75$, by definition "per-capita" scaling coefficient is 0.75 − 1 = −0.25). Unfortunately, existing scenario studies[45,46] stop at country-level averages and cannot tell us how many people will live in very-high- or very-low-density neighborhoods. For example, in one potential scenario, UN-Habitat projections[45] suggest that average urban density in low-income countries could increase to 164% of current levels by 2050. If this happens, and we apply our scaling exponent ($\delta = 0.75$), the per-capita built mass in a neighborhood would decrease by 12% ($1.64^{-0.25}$=0.88) compared to linear projections. However, this average-density scenario still overlooks the heterogeneous non-linear effects across various neighborhoods. Therefore, future projections must shift from average densities to distributions among multiple density-levels if we are to understand the real material stakes of urban growth.

Beyond material efficiency, urban scale and density also lead to significant socioeconomic advantage. Large cities enable deeper labor market pooling, specialized supplier networks, and extensive knowledge-sharing, all of which bolster innovation and productivity[47]. For instance, a bigger urban population allows a finer division of labor and supports niche industries and services that require a broad customer base, which tend to cluster in major cities[48]. In fact, even a tiny fraction of people with a niche interest can form a critical mass in a metropolis[49]. High population density further reduces the spatial separation between people and opportunities, facilitating efficient matching in the labor market and easy access to specialized services[50]. Dense neighborhoods also foster more frequent face-to-face interactions and serendipitous encounters—the "unplanned synergies" that have been widely recognized in urban studies[51–53] as drivers of innovation and social connectivity. Together, these mechanisms suggest that urban scale and density are key drivers of social and economic opportunities, providing access to spatialized markets, services, and social networks that less dense or smaller urban areas cannot easily support.

While increasing city sizes and neighborhood densities can enhance material efficiency and bring socioeconomic advantages, unchecked growth and densification can intensify social and public health challenges. For instance, larger cities often experience disproportionate increases in congestion, crime, infectious diseases, as these outcomes scale superlinearly with population size, reflecting the complex dynamics of urban environments[54–56]. Moreover, in densely populated urban slums, shared sanitation facilities are prevalent due to limited space and resources, but poor maintenance and hygiene often lead to health risks and diminished quality of life[57]. The collective responsibility for maintaining such facilities frequently results in neglect, as individuals are less motivated to care for shared spaces. These challenges highlight the need to balance material efficiency with investments in infrastructure that ensure adequate living standards[58] and public health in high-density urban areas.

Taken together, our findings collectively underscore the importance to recognize these linked, multi-scale material implications as intrinsic properties of cities. Viewing cities as complex adaptive systems—whose performance depends jointly on their overall size and their hierarchically nested internal structure[59,60] will enable more holistic and effective policy interventions in the pursuit of a more productive, equitable, and sustainable urban future.

## Methods

### Historical trend of global built structure mass

We derived historical global population data from our world in data (https://ourworldindata.org/grapher/population). We acquired the global built structure mass and global biomass dataset from Elhacham et al. (2020) from 1900 to 2015 and projection data from 2015 to 2037[1]. We used linear interpolation (zoo::na.approx()) to interpolate missing global biomass value during 1991-1999, 2001-2009, 2011, 2013-2016. To calculate the built structure to human mass ratio, we estimated weighted average human body mass as approximately 50 kg. We derived this coarse-grained estimation by combining the global average age distribution (25% 0-14, 65% 15-64, 10% 65+ in year 2023 based on World Bank data, https://data.worldbank.org/indicator/SP.POP.0014.TO.ZS) with the age-specific body mass (25kg for 0-14, 60kg for 15-64, 50kg for 65+). We are aware that age-specific body mass varies across countries, regions, diets, and a range of other factors, however, the qualitative conclusions presented in our work are not sensitive to the uncertainties associated with this estimation.

### Vegetation biomass

We extracted city-specific vegetation biomass from a recently published global map of vegetation carbon that includes both aboveground and belowground biomass carbon[5]. The vegetation carbon data is based on 2010. We then converted biomass carbon into dry-weight vegetation biomass using a convection coefficient of 0.48g carbon per gram of dry-weight biomass[61]. We extracted city-specific vegetation biomass based on city definitions described in the following section.

### Definition of cities

In this study, we define cities using the concept of Urban Cores (UC) provided by the Global Human Settlement Layer (GHSL). GHSL, developed by the European Commission's Joint Research Centre, delineates urban cores based on continuous areas of built-up land and population density, derived from satellite imagery and census data[21]. An urban core is typically characterized by contiguous built-up surfaces with high population density, representing the primary agglomeration of urban activity within a region. Following established conventions, we include only cities with a minimum population size of 50,000 residents in our analysis, ensuring consistency and comparability with other global urban studies [4,12]. This operational definition allows us to systematically evaluate and compare urban built environments and associated resource demands across a diverse range of global cities. Our main finding that economies of scale in urban built stock is prevalent in cities worldwide is robust to the arbitrary cutoff sizes of cities. In fact, we would argue that the exact cutoff of city definition is not so important given the self-similarity, or fractal-like feature of urban built stock.

### Definition of neighborhoods

We delineate neighborhoods with the H3 global hexagonal grid (developed by Uber)[32]. Hexagons have two properties that are critical for our analysis. First, among the few regular polygons that tessellate the plane, hexagons minimize the perimeter-to-area ratio, yielding compact, near-circular units that reduce edge effects when we aggregate spatial data. Second, hexagons tile space without the directional bias that characterizes squares (Manhattan effect) or triangles, thereby supporting isotropic distance measures that better approximate real‑world movement patterns. These features mirror the intuition of Christaller's central‑place theory[62], which idealizes market areas as hexagons to maximize coverage while minimizing overlap. Here we use the hexagons at resolution of level-6, where each cell has an average edge length of ~3.7 km and an area of 37 km², a scale that captures the extent of day‑to‑day human activity: the cell radius can be covered in ~15 min by bicycle (15 km/h) or ~45 min on foot (5 km/h). The resulting neighborhoods are therefore large enough to encompass the functional mix of a self‑contained urban district, yet fine‑grained enough to retain intra‑city variation in the variables we study.

**Bottom-up approach of urban built stock**
To quantify the material stocks embedded in the built environment of cities, we devised a workflow that is the first global analysis that builds up the total mass of urban buildings and paved surfaces from the ground up. This method separately estimates the volumes of building and areas of mobility infrastructure within each city boundary and relies on well-established, type-specific and region-specific material intensity factors to convert built volume, road surface, and pavements into mass, detailed below.

**1) Stock of buildings.** We use raster-based estimates of building volume provided independently by Esch et al. (2022)[18], Li et al. (2022)[16], and Liu et al. (2024)[17], which offer 3D characterizations of urban form at various spatial resolutions (Supplementary Table.1). To minimize dataset-specific bias, we calculate the average building volume within each city boundary using all three datasets.

We then converted building volume to built stock mass using a detailed material intensity framework that considers both (1) building types and (2) regional construction practices developed by Haberl et al. (2025) for analyzing global buildings [23]. We distinguish building types based on GHS-BUILT-S classifications, as the distinct function and form of different building types demands different building standards and consequently different material intensities independent of country and region. Similarly, material intensity is also a function of country and region due to the dependence of building on local climate, building practices, developmental level, etc. In Supplementary Table.2 we provide a summary look-up table for matching building types to material intensities. For the detailed regional breakdown (OECD, North America, Japan, China, etc.), please refer to Haberl et al. (2025) Extended Table.5.

**2) Stock of mobility infrastructure and other pavement.** For impervious surface data, we use the GISA-10m product from Huang et al. (2022)[63], which maps global impervious cover at 10-meter resolution. To isolate the portion of impervious surface attributable to pavement (*e.g.*, roads, sidewalks), we subtract the building footprint area—estimated from the previous step by averaging Li et al. (2022)[16], Liu et al. (2024)[17] and Esch et al. (2022)[18] —from the total impervious surface within each city.

After extracting pavement area, we further tease out 5 grades of roads network vs. other pavements (sidewalks, parking lot) given the big difference in their required structural intensity and consequently material intensity. For each city, we calculate the total length and total area of the road network per road grade: with decreasing material intensity, highway, primary, secondary, tertiary, and local. We then convert road area to road stock mass based on a type-specific and region-specific material intensity lookup table (summarised in Supplementary Table. 3, full table uploaded as separate supplementary file, derived from ref [22])

For other pavements (*e.g.*, sidewalks, parking lot), we assume a material intensity of 570 kg/m² (equivalent to about 20 cm of concrete or asphalt), a value lower than typical road material intensity adopted by previous meta-analysis in the field of industrial ecology (see Frantz et al. 2023 Extended Data Table.15 for parking lots[13]).

The total mass of urban built stock is the sum of the estimated building, road mass and other pavement mass. This bottom-up approach provides a consistent and scalable method to compare material stocks across global cities using spatially explicit data.

**Data acquisition for urban population**
We used the WorldPop dataset to calculate the total population of each city by summing pixels within city boundaries from WorldPop raster datasets. We chose WorldPop due to its high spatial resolution (100m), detailed demographic representation, frequent updates, and open accessibility, providing reliable population data that better captures population distributions compared to other gridded population datasets[64,65]. WorldPop integrates diverse data sources, including satellite imagery, census, and ancillary data, making it particularly suitable for accurately mapping populations within urban areas.

**Scaling theory and fixed slope analysis**
Scaling theory studies the change of system properties as a system changes in its size[66]. Its conceptual foundation can at least be traced back to early insight regarding proportional scaling up (*i.e.*, linear scaling) built infrastructure in Galileo's *Two New Sciences*[67]. Galileo deducted that scaling structures proportionally in size increases their volume and mass at a faster rate than the increase of strength, leading to structural failures. Mathematically, this type of relationship is expressed through power-law equations of the general form:

$$Y = Y_0 N^\beta$$

In this equation, *Y* denotes a measurable property or characteristic of the system, *N* represents system size, *β* is the scaling exponent capturing how quickly a property changes relative to system size, and $Y_0$ is the normalization constant, representing the elevation of the scaling relationship on a log-log plot[68,69], providing insights into the baseline or developmental level of the system (*e.g.*, a collection of buildings, a taxon of multiple species, or a group of cities).

Aside from its widespread application in physics and engineering, scaling theory significantly influenced biology through the early 20th-century subfield of allometric analysis, enhancing the adoption and interpretation of Darwin's evolutionary theory[70–73]. Notable examples include scaling relationships such as organism body mass with metabolic rate and tree height with stem diameter. These foundational biological concepts were further expanded by West, Brown, and Enquist, who formalized biological scaling laws, laying the groundwork for applications in other complex systems, including urban contexts initiated by Bettencourt and colleagues[26].

Scaling theory has recently been widely adopted to enhance our understanding of urban systems, elucidating how a wide range of urban indicators (wealth, innovation, crime, infrastructure, waste, etc.) scale with city population size[12,74]. Variations in the exponent indicate critical nonlinearities—economies or diseconomies of scale—that are often obscured by simple per-capita measures. Superlinear scaling (*β>1*) characterizes intensified social interactions and innovation typically seen in larger cities, whereas sublinear scaling (*β<1*) describes constraints related to physical infrastructure and spatial limits[10].

To statistically manage correlations between the scaling exponent *β* and normalization constant $Y_0$, the method of fixing slope and estimating elevation is commonly applied, originating from biological analyses[68,75,76]. We use this statistical approach to derive nation-specific baselines in cross-city analyses and city-specific baselines at neighborhood scales, allowing us to normalize data effectively and focus explicitly on scaling slopes to reveal fundamental urban dynamics.

**Simulation of the emergence of city-level scaling from neighborhood-level disparity**
We developed a simulation framework to investigate how variations in the neighborhood-level Zipf's law distribution of population (characterized by the Zipf shape parameter *s*) influence city-level scaling patterns, given a fixed neighborhood-level scaling exponent (*δ* = 0.75). The total built mass ($M$) of cities was modeled as the sum of neighborhood-level built masses ($m_i$), where $m_i = n_i^\delta$ and $n_i$ is the population of the $i$-th neighborhood. With these simulated cities, we then estimated the scaling coefficient ($\hat{\beta}$) relating $M$ to $N$ using log-log regression.

To explore the emergent relationship between *s* and $\beta$, we performed a parameter sweep of *s* values and analyzed the city-level scaling patterns across a range of scenarios. This approach revealed how the aggregation of neighborhood-level power-law distributions affects the observed scaling exponent at the city level. To examine the uncertainties of $\beta$, we introduced two random processes: 1) we drew a random s parameter from a normal distribution, centered at the given s with the standard deviation (0.59) estimated from the observed distribution (Fig.3a inset; Fig.4a); 2) we grouped the cities randomly into "countries" by Dirichlet distribution with shape parameter of 0.1 to generate a power-law like distributions of countries ranked by numbers of cities. By applying these two random processes, we produce a distribution of theoretical β values.

**Derivation city-level scaling law from neighborhood-level disparity**

We partition each city into *K* standard hexagonal neighborhoods, labeling them so that $n_{(1)}$ is the largest neighborhood population, $n_{(2)}$ the second largest, and so on and so forth. Likewise, $m_{(1)}$ is the largest neighborhood built mass, $m_{(2)}$ the second largest, etc. Empirical evidence suggests these quantities follow Zipf-like rank–size laws:

$$n_{(i)} \sim i^{-s_n}$$

and

$$m_{(i)} \sim i^{-s_m}$$

where the negative exponents $s_n$ and $s_m$ indicate that population and built mass decrease with increasing rank *i*. If we posit that the same neighborhood that ranks *i*-th in population on average also ranks *i*-th in built mass, we connect these scalings to obtain:

$$m_{(i)} \sim i^{-s_m} = \left(i^{-s_n}\right)^{\frac{s_m}{s_n}} = \left(n_{(i)}\right)^{\frac{s_m}{s_n}}$$

Defining a new parameter

$$\delta = \frac{s_m}{s_n}$$

$\delta$ is thus the neighborhood-level scaling exponent relating built mass to population:

$$m_{(i)} \sim n_i^{\delta}$$

If $s_m < s_n$, then $\delta < 1$, implying sublinear growth of $m_{(i)}$ with $n_{(i)}$.

Note that in this rank-order formulation, $s_n$ is the neighborhood version of Zipf's city rank-order parameter[77]. Empirical evidence suggests that city-level *s* are frequently found very close to 1 (0.8-1.2). Based on our analysis, the neighborhood level data (Fig.4a) has a Zipf parameter $s_n$ in the range between 0.7 (25% quantile) to 1.4 (75% quantile).

In our dataset, $s_m$ is consistently smaller than $s_n$, indicating neighborhood-level mass is more homogenous (in the limit case, *s* = 0 represents all neighborhoods are exactly equal in size, the higher the *s*, the more disparate) than neighborhood-level population distribution.

We then define the total city population

$$N = \sum_{i=1}^{K} n_i$$

and total city built mass

$$M = \sum_{i=1}^{K} m_i,$$

and write

$$M \sim N^\beta,$$

where $\beta$ is the city-level scaling exponent. If $s_n \gg 1$, the largest population $n_{(1)}$ dominates $N$, and similarly if $s_m \gg 1$ the largest built mass $m_{(1)}$ dominates $M$. Then

$$N \approx n_{(1)}, \quad M \approx m_{(1)} \sim (n_{(1)})^\delta,$$

so

$$M \sim N^\delta \implies \delta = \beta.$$

Conversely, if $s_n$ is close to zero, so that all neighborhoods are similar in size, then we typically have $N \sim K$. If $s_m$ is close to zero as well, we will have $M \sim K$ too, giving:

$$M \sim N \implies \beta = 1.$$

Hence, $\beta$ are bounded between $\delta$ (in a largest-neighborhood–dominated regime) and 1 (in a homogenous regime).

**Statistical analyses**
We used linear regression to analyze the match between city mass and urban population (Fig.2ab, Supplementary Figs.2,3; lm, R package 'stats', v.4.4.0). Both city mass and urban population are right skewed and were log10 transformed before regression analysis, consistent with logarithmic scale presentation of our figure axis. For Fig.2b, we first calculated the country-specific intercepts using linear mixed effects model (R package "lme4", v1.1-36) treating cities as nested inside of countries with fixed slopes. We used the resulting country-specific intercept to normalize all city data (bring city groups to the same country-specific intercept) before analyzing the general relationship between city mass and population size (Fig.2b). In our multi‑level fixed–effect scaling model (Fig.3; package "lme4" v.1.1-36), we regress the logarithm of built mass $M_{ijk}$ in neighborhood $k$ of city $j$ in country $i$ on the logarithm of neighborhood population $N_{ijk}$, while absorbing separate intercepts for each country and each city. In compact form, the model is

$$log(M_{ijk}) = \gamma_0 + \gamma_i + \gamma_j + \delta \cdot log(N_{ijk}) + \varepsilon_{ijk}$$

where $\gamma_0$ is the global baseline intercept, $\gamma_i$ the deviation for country $i$, and $\gamma_j$ the deviation for city $j$ (each city $j$ being nested within its country $i$). The slope $\delta$ captures the universal elasticity of built mass with respect to population across all neighborhoods, and $\varepsilon_{ijk}$ is the neighborhood‑level error term.

Identification is achieved by constraining $\sum_i \gamma_i = 0$ and $\sum_{j \in i} \gamma_j = 0$, so that $\gamma_0$ remains the grand intercept on the log scale. We estimate this high‑dimensional fixed–effect regression via dummy‑variable (within) transformation: first subtracting country and city means to partial out $\gamma_i$ and $\gamma_j$, then fitting $\delta$ by ordinary least squares on the demeaned log–population and log–mass. Standard errors are clustered at the city level to allow for arbitrary spatial correlation within each city. Because the outcome and predictor are both in logs, the estimated $\hat{\delta} < 1$ directly quantifies sublinearity: a 1 % increase in local population corresponds to only a $\hat{\delta}$ % increase in built mass, uniformly across all countries and cities. We further validate robustness by fitting the analogous mixed‑effects model with random intercepts $u_i$ and $u_j$, obtaining virtually identical estimates of $\delta$. All analyses were conducted in R (v.4.4.0)

**Data availability**
Data will be made available on public online depository (figshare) upon publication.

**Code availability**
Code will be made available on public online depository (figshare) upon publication.

**Acknowledgement**
This work was supported by a New York University internal grant awarded to M.L. and K.H. We thank members of Lu lab (Aiyu Zheng, Jacob Levine, Ignacio Arroyo), Karen Seto and members of Seto lab, Jose Lobo,  and colleagues at Department of Environmental Studies for helpful comments on the manuscript.

**Author Contribution**
K.H. and M.L. developed the overall conceptual framework, carried out analysis, designed visualization, wrote the paper together. Both authors contributed equally.

**Competing Interests Statement**
The authors declare no competing interests.

**Supplementary Information**

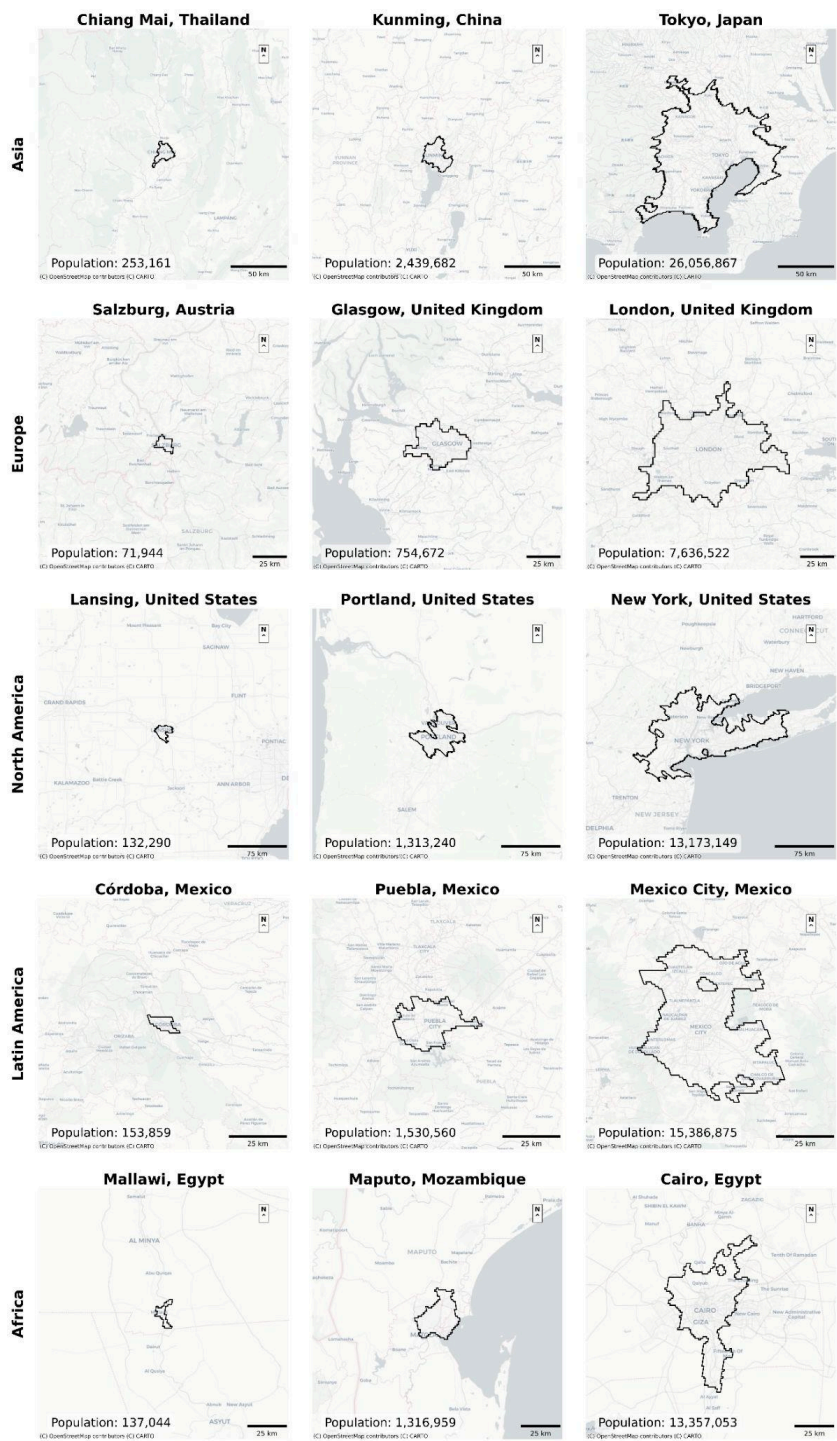

**Supplementary Figure 1| Illustration of city definition, example cities across various continents demonstrating different scales of cities.**

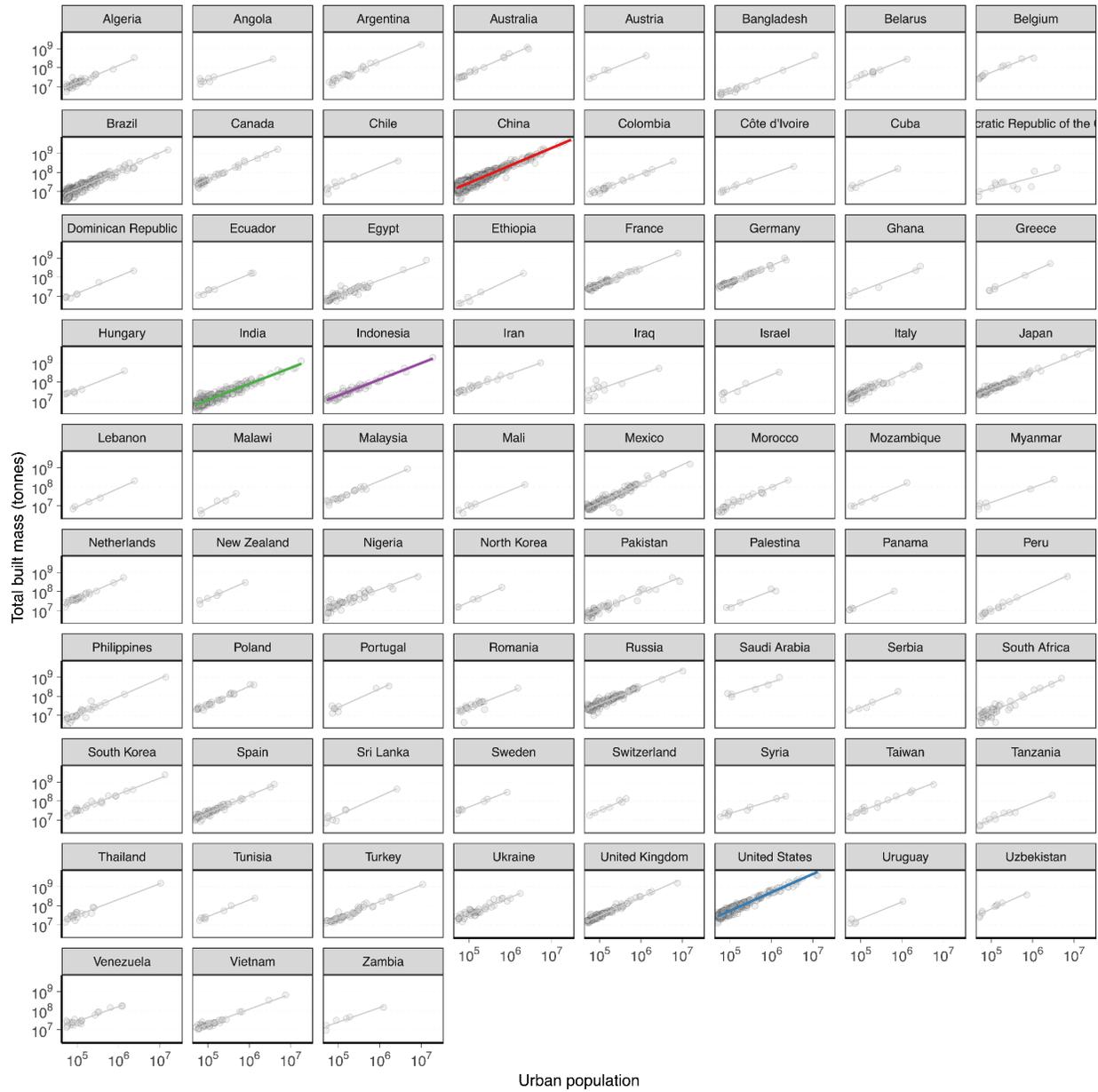

**Supplementary Figure 2| Sublinear urban scaling of built mass and variations across all countries.** We have 74 countries meeting our criteria of inclusion (n_cities>5).

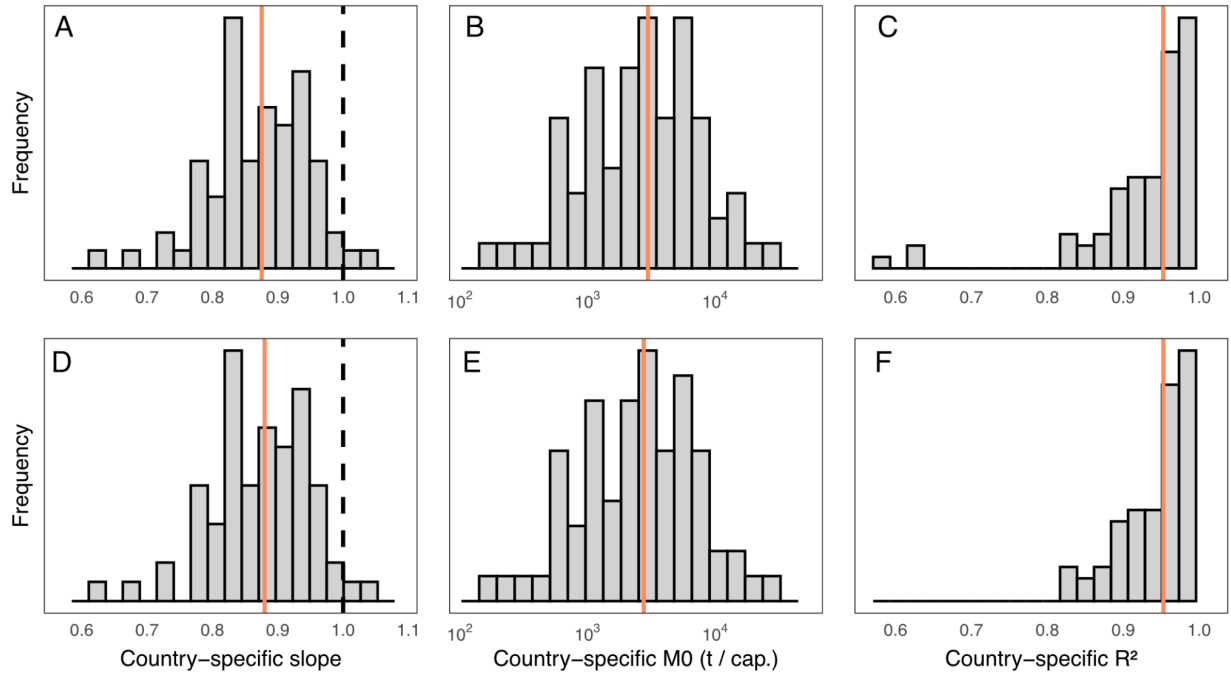

**Supplementary Figure 3| Country-level scaling analysis stats.** The top row (A,B,C) includes all 74 countries with more than 5 cities, while the second row (D,E,F) excludes 3 countries with R2 lower than 70%. The overall country-level stats are not sensitive to the inclusion of these 3 countries with relatively low goodness of fit.

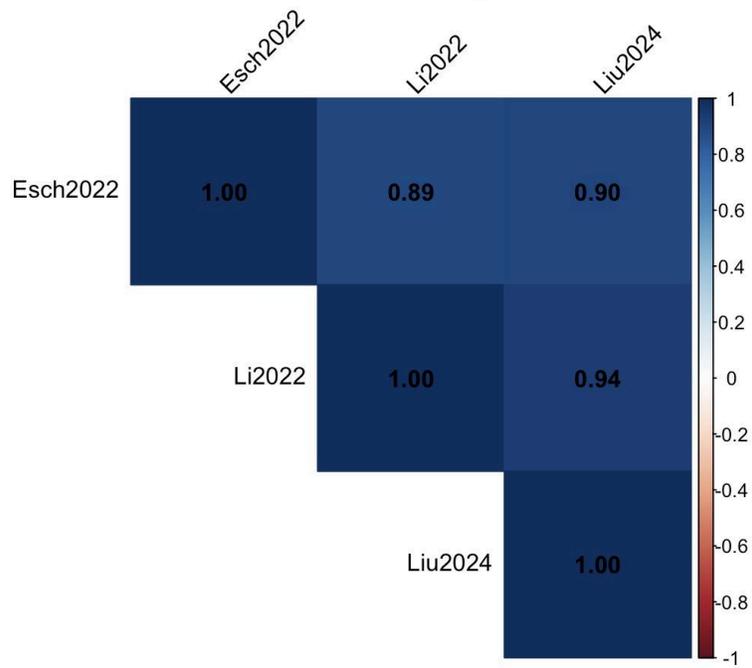

**Supplementary Figure 4| Correlation matrix of building mass estimates across three global datasets.** The heatmap shows Pearson correlation coefficients between city-level building mass estimates from Esch2022, Li2022, and Liu2024 datasets. Analysis includes cities with complete data across all three datasets ($n$=3721).

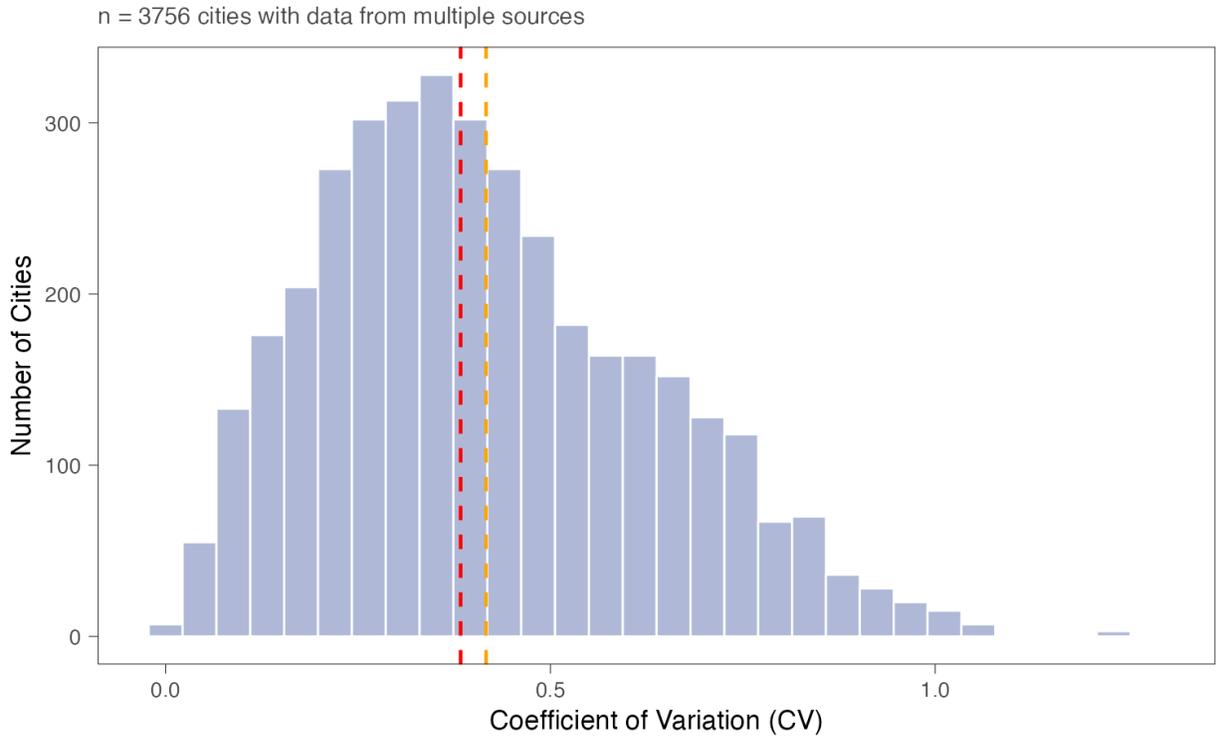

**Supplementary Figure 5| Distribution of coefficient of variation (CV) of urban building mass estimates across cities from multiple data sources.** This histogram illustrates the distribution of the coefficient of variation (CV) of urban building mass estimates across cities, calculated from three global building volume datasets: Esch et al. (2022), Li et al. (2022), and Liu et al. (2024). The CV is derived as the ratio of standard deviation to mean building mass for each city, quantifying the relative variability between different data sources. Only cities with data available from at least two sources are included in the analysis (Zahedan, Iran is the only city with one source and thus excluded from variability analysis). The red dashed line indicates the median CV, while the orange dashed line represents the mean CV.

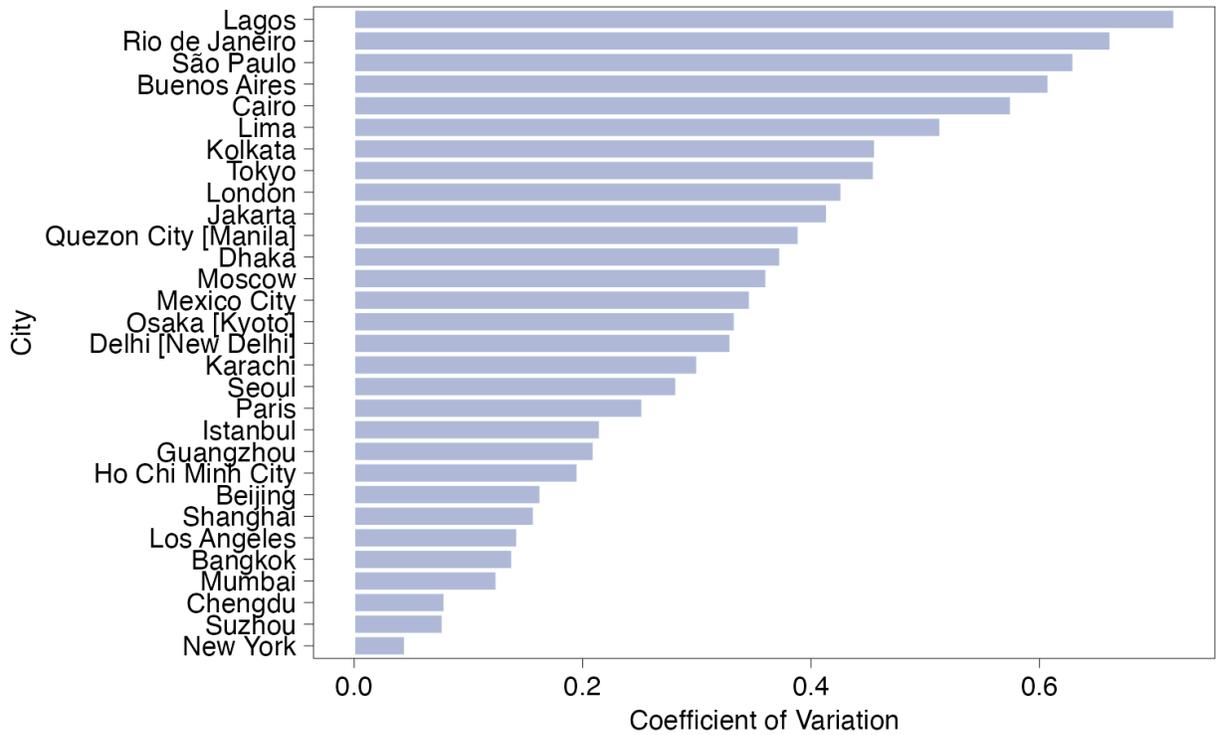

**Supplementary Figure 6| Coefficient of variation in urban building mass estimates for the 30 most populous cities with multi-source data.** Cities are ranked by CV magnitude, showing the relative variability in building mass estimates between the three global datasets (Esch et al. 2022, Li et al. 2022, and Liu et al. 2024). Higher values indicate greater disagreement between data sources for that particular city. Cities are selected based on 2015 population data and represent the largest urban areas where building mass estimates are available from multiple sources.

**Supplementary Table 1| Summaries of input 3D building datasets**

| Source | Input Imagery (years) | Resolution | RMSE (volume) |
|---|---|---|---|
| Esch et al. (2022) | TanDEM-X amplitude (2011, 2013); WSF 2019 mask (Sentinel-1/2 pre-2019); | 90 m | 1.54 m³/m² |
| Li et al. (2022) | Landsat-8 (2015); Sentinel-1 (2015–2016); | 1 km | 0.619 m³/m² |
| Liu et al. (2024) | Sentinel-1 (2015); ALOS (2015); Landsat-8 (2014-2016); | 500 m | 0.243 m³/m² |

**Supplementary Table 2| Material intensity as a function of building type and region.**

| Building type | Building height (m) | GHS-BUILT-S (class) | Region-specific material intensity (kg/m$^3$) |
|---|---|---|---|
| Lightweight (LW) | < 3 | Residential + non-residential | 151–154 |
| Residential single-family house (RS) | 3–12 | Residential | 125–526 |
| Residential multi-family house (RM) | 12–50 | Residential | 315–662 |
| Non-residential (NR) | 3–50 | Non-residential | 273–654 |
| High-rise (HR) | 50–100 | Residential + non-residential | 312–330 |

**Supplementary Table 3| Material intensity of roads as a function of road type, region and climate.**

| Road Type | # Lanes (approx.) | Lane width (per lane, m) | Region- and climate-specific material intensity (kg/m²) |
|---|---|---|---|
| Highway | 3 – 7 | 3.0 – 4.0 | 910.6 – 2883.5 |
| Primary | 2 – 5 | 3.0 – 4.0 | 803.9 – 2331.5 |
| Secondary | 1 – 5 | 2.4 – 4.1 | 768.1 – 1817.9 |
| Tertiary | 1 – 4 | 2.4 – 4.1 | 647.7 – 1750.8 |
| Local | 1 – 3 | 2.4 – 4.1 | 538.5 – 1548.2 |